\documentclass[a4paper]{jpconf}
\usepackage{graphicx}
\usepackage{blindtext, graphicx}
\usepackage{cite}
\usepackage{hyperref}
\usepackage{amssymb}
\usepackage{float}
\usepackage{listings}
\usepackage{afterpage}
\usepackage{subfig}

\begin{document}

\title{Deep Neural Networks for Physics Analysis on low-level whole-detector data at the LHC}
\author{Wahid Bhimji$^1$, Steven Andrew Farrell$^1$, Thorsten Kurth$^1$, Michela Paganini$^{1,2}$, Prabhat$^1$, Evan Racah$^1$}
\address{$^1$ Lawrence Berkeley National Laboratory, 
Berkeley, CA 94720 USA}
\address{$^2$ Department of Physics, Yale University, New Haven, CT 06520, USA}
\ead{wbhimji@lbl.gov}

\begin{abstract}
There has been considerable recent activity applying deep convolutional neural nets (CNNs) to data from particle physics experiments. Current approaches on ATLAS/CMS have largely focussed on a subset of the calorimeter, and for identifying objects or particular particle types. We explore approaches that use the entire calorimeter, combined with track information, for directly conducting physics analyses: i.e. classifying events as known-physics background or new-physics signals.

We use an existing RPV-Supersymmetry analysis as a case study and explore CNNs on multi-channel, high-resolution sparse images: applied on GPU and multi-node CPU architectures (including Knights Landing (KNL) Xeon Phi nodes) on the Cori supercomputer at NERSC.

We compare statistical performance of our approaches with selections on high-level physics variables from the current physics analyses, and shallow classifiers trained on those variables. We also compare time-to-solution performance of CPU (scaling to multiple KNL nodes) and GPU implementations.
\end{abstract}

\section{Introduction}

A major aim of experimental high-energy physics (HEP) is to find rare signals of new particles produced in large numbers of collisions at accelerators such as the Large Hadron Collider (LHC) and the ATLAS and CMS experiments. Improvements in classifying these collisions would improve sensitivity to discoveries that could overturn our understanding of the universe at the most fundamental level. 
Neural Networks have been used in high energy physics for some time. Recently attention has focused on deep learning to improve sensitivity, as well as tackle the increase in detector resolutions and data rates in HEP experiments. 

Here we evaluate the use of deep Neural Network's (NN) on low-level detector data directly for physics analysis without reconstruction of physics objects like jets; without tuning of analysis variables; and using data from the entire calorimeter together with other sub-detectors. We use modern deep learning methods for performance; relate performance to that of physics variables; and ensure results are robust to factors such as pileup and alternative physics models. 

The second goal of this work is to ensure these NN architectures run efficiently with popular deep-learning software frameworks on the Cori supercomputer at the National Energy Research Scientific Computing Center (NERSC). This is a major computing resource that is predominately composed of Intel Knights Landing (KNL) XeonPhi CPUs. We seek to improve single node performance on these CPUs, providing timings, optimisations and recipes, as well as distributed training up to the full scale of the machine (10k KNL nodes).

We provide background, in Section \ref{sec:physics}, on the physics use-case studied and benchmark selections; in Section \ref{sec:CNN}, the CNN architecture we employ; and, in Section \ref{sec:NERSC} the computing resources at NERSC. We then provide physics performance results in Section \ref{sec:Results} together with interpreting how the learned network compares to the physics variables used in the benchmark analysis (in Section \ref{sec:Results2}). Then in Section \ref{sec:timing} we show timing results for implementations in various frameworks and on GPU and CPU resources at NERSC, before concluding in Section \ref{sec:conclusion}.

\section{Background}
\subsection{Physics and datasets}
\label{sec:physics} 

For this study, we take as a use-case a particular analysis searching for new massive supersymmetric (`RPV-Susy') particles in multi-jet final states at the LHC \cite{ATLAS:2016nij}. We focus on `gluino-cascade decays' with gluino and neutralino masses of 1400 GeV and 850 GeV by default (and varied for performance studies). 

We use the Pythia event generator \cite {Pythia} interfaced to the Delphes fast detector simulation \cite{Delphes} and using the default Delphes ATLAS detector configuration (`card'). We generate events for two classes: the RPV-Susy signal and the most prevalent background (`QCD'). In order to have sufficient quantities of background events at high transverse momentum ($p_T$) that are harder to separate, the QCD sample is generated in ranges of $p_T$ and then weighted according to the amount expected (cross-section) in both training loss and the evaluation.

Before training our network we apply some of the physics selections used in \cite{ATLAS:2016nij} to filter images to those more challenging to discriminate, resulting in a training sample of around 400k events. We compare the performance of our deep network to our own implementation of the full selections in \cite{ATLAS:2016nij} as a baseline benchmark. To produce the jet variables we use the same jet algorithm as in the physics analysis with parameters given in \ref{tab:presel} and applied using `FastJet' \cite{fastjet} within Delphes. The preselection and benchmark selections are also summarized in \ref{tab:presel}. We have verified that our samples and baseline selections give performance comparable to that obtained in \cite{ATLAS:2016nij} providing a meaningful benchmark even though those selections were not tuned for exactly these datasets. We also compare our analysis to shallow classifiers based on these physics variables.  As inputs to those classifiers, we use the full four-momentum of five jets with highest transverse momentum as well the jet variables used in the physics analysis selections: namely the sum of jet mass, number of jets, and $\Delta\eta$ between the leading 2 jets. 

Data from the surface of the cylindrical detector 
 is represented as a 2D image with coordinates corresponding to azimuthal angle $\phi$ and pseudorapidity $\eta$. For the pixel intensity in this image, we use either the overall energy deposited in the combined calorimeter or split the energy deposited in the electromagnetic and hadronic calorimeters, and add the number of tracks formed from the inner detector in that region to provide a three channel image. This is similar to the approach of  \cite{jetimage1}\cite{jetimage2} except that we use large images covering the entire detector, and use these directly for classifying entire events rather than individual objects.  We choose to bin the energy (and number of tracks) into uniform 64x64 bins,which correspond to the approximately 0.1x0.1 ($\eta$ x $\phi$) resolution of the ATLAS hadronic calorimeter, but also use larger 224x224 images for large-scale studies that could capture the granularity of the ATLAS electromagnetic calorimeter and future detectors.

\subsection{Neural Net Architecture}
\label{sec:CNN} 

\begin{figure*}[!t]
\centering
\includegraphics[width=0.80\textwidth,trim={0 3.9cm 0 0} ,clip]{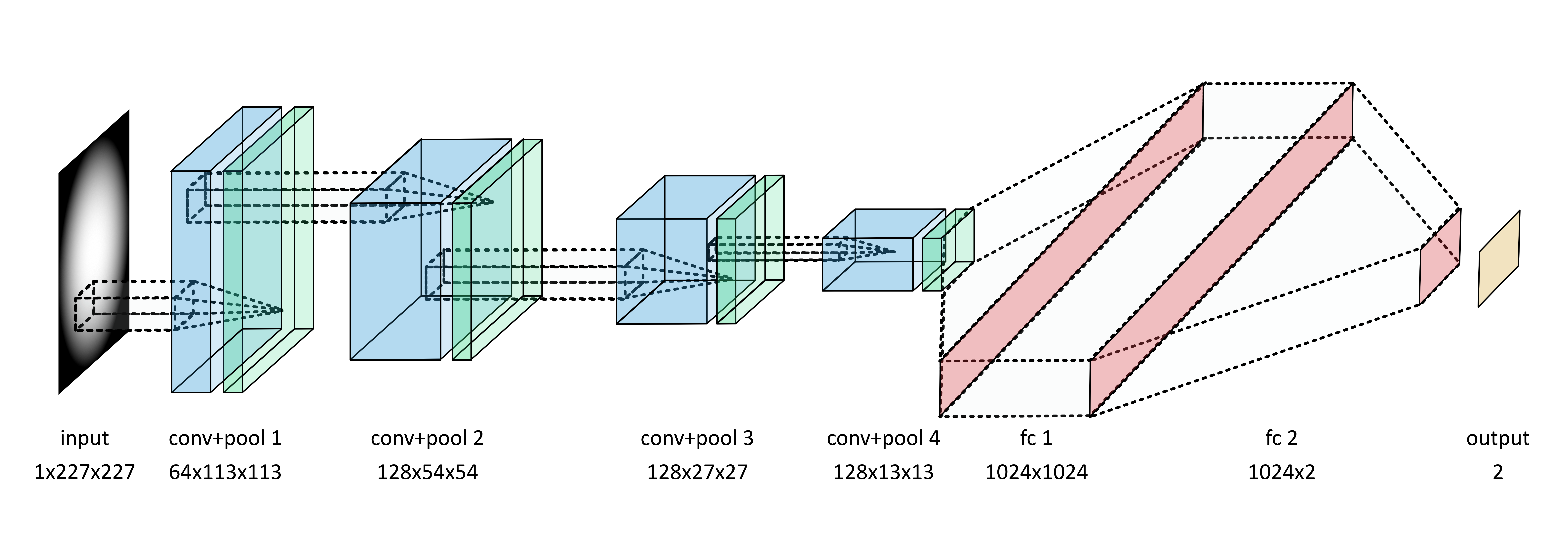}
\caption{CNN architecture}
\label{fig:CNN}
\end{figure*}

We perform binary classification on the resulting images using a Convolutional Neural Net (CNN) \cite{lecun1998gradient} comprised of 3 convolution+pooling units (or 4 for larger images) with rectified linear unit (ReLU) activation functions \cite{nair2010rectified}. 
These layers output into two fully connected layers which project into a two-dimensional vector on which a softmax function is applied to determine the signal and background class probabilities. We use softmax with cross-entropy as the loss function and the ADAM optimizer \cite{DBLP:journals/corr/KingmaB14}. The network employed is shown in figure \ref{fig:CNN}. 


\subsection{Computational Resources}
\label{sec:NERSC} 

The CPU timing and scaling results reported here used the NERSC Cori system: a Cray XC40 with 2388 Intel (E5-2698 v3) Haswell and 9688 Intel XeonPhi 7250 Knight's Landing (KNL) compute nodes. Each KNL processor includes 68  1.4GHz cores with 4 HyperThreads and peak performance of 6 TeraFLOP/s (single-precision). Each Haswell node has dual 16-core 2.3 GHz processors with peak performance of 2.4 TeraFLOP/s. We compare performance to a standalone machine containing Titan X (Pascal) GPUs each capable of 10.2 TeraFLOP/s peak. 

\section{Results}

\begin{figure*}[!t]
\centering
\begin{minipage}{.48\textwidth}
\includegraphics[width=\textwidth]{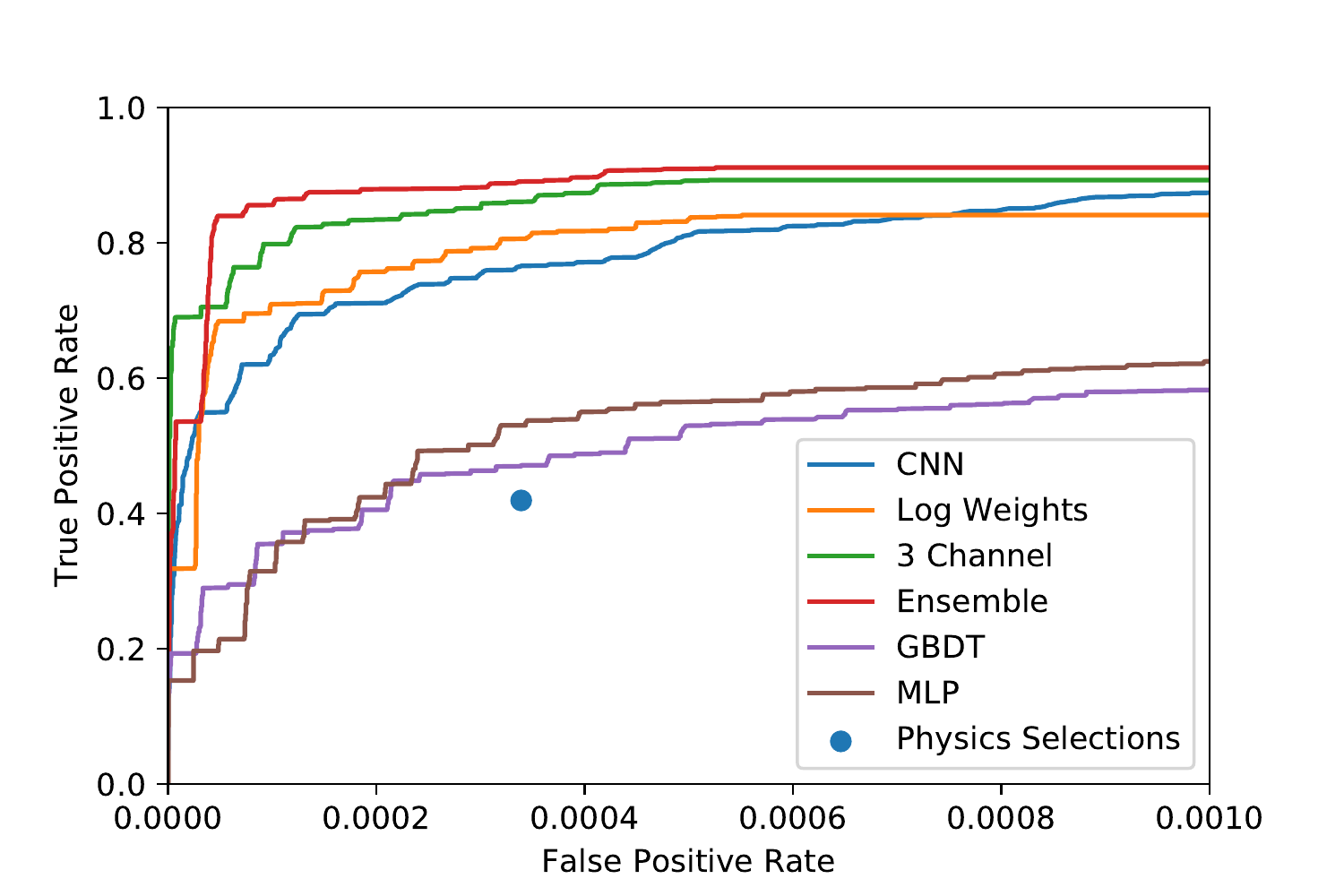}
\caption{ROC curve comparison of different CNN implementations with physics selections and shallow classifiers}
\label{fig:roc1}
\end{minipage}
\hfill
\begin{minipage}{.48\textwidth}
\includegraphics[width=\textwidth]{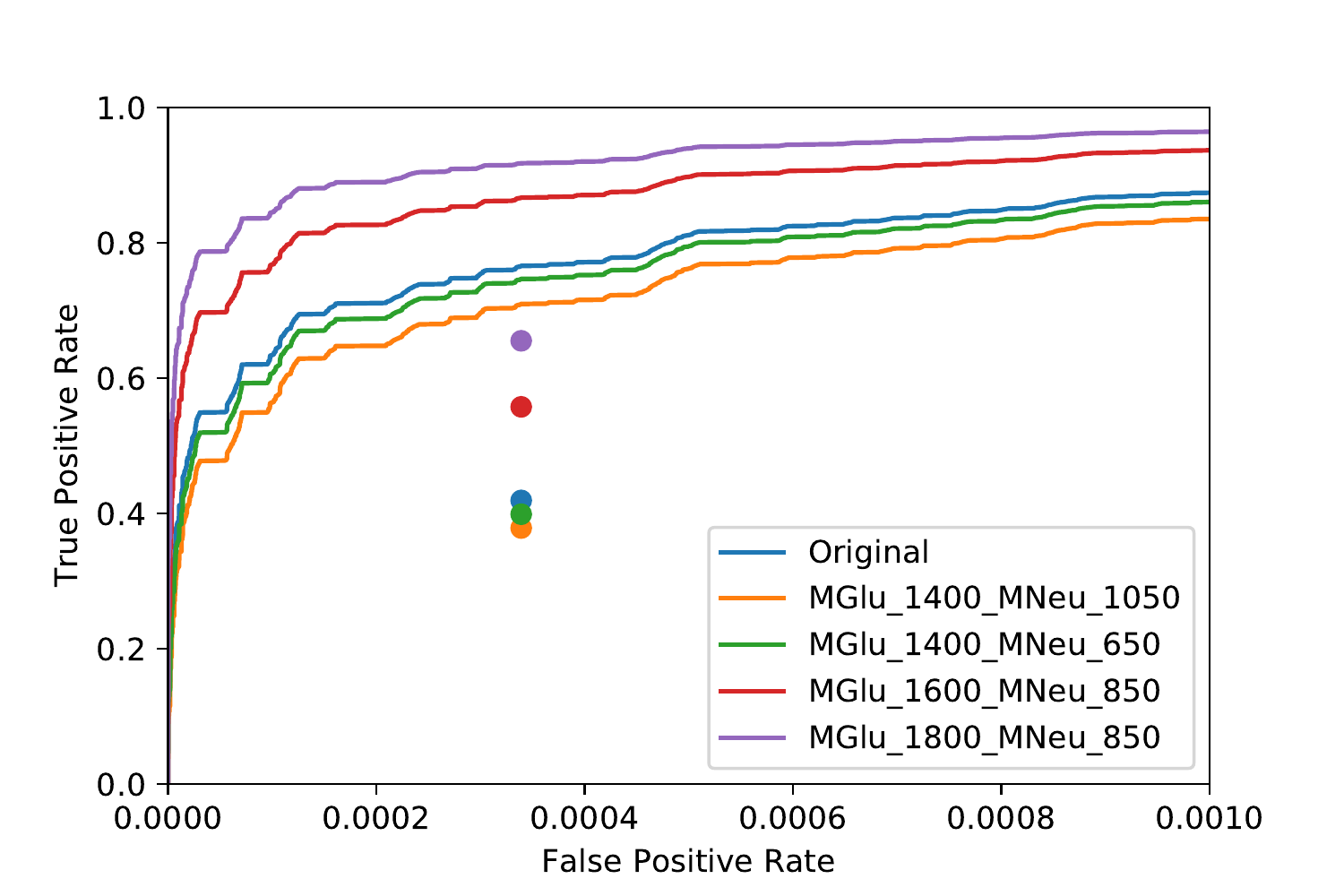}
\caption{ROC curves for application of CNN to different signal mass samples}
\label{fig:roc4}
\end{minipage}
\end{figure*}

\subsection{CNN Performance}
\label{sec:Results}
In order to obtain good significance to new physics it is necessary to combine high signal efficiency (True Positive Rate (TPR)) with very high background rejection (low False Positive Rate (FPR)). In figure \ref{fig:roc1} we show the ROC curve of TPR vs FPR for the CNN architecture described in \ref{sec:CNN} with a single channel for calorimeter energy (labelled `CNN'). We compare to the physics selections described in \ref{sec:physics}. We achieve increased signal efficiency at same background rejection without using physics-based jet variables. We also compare the AMS (approximate median significance) \cite{adam2014learning} which gives a single measure accounting for initial pre-selection efficiency and relative signal/background cross-sections, and achieve a 1.8x improvement in AMS at the same background rejection point. 

\emph{\bf Comparison to shallow classifiers:} We  also compare to shallow machine-learning approaches using high-level physics variables as inputs as detailed in Section \ref{sec:physics}. The performance for a gradient boosted decision tree (GBDT) and a 1 hidden layer fully-connected NN (MLP) are shown in figure \ref{fig:roc1}: these outperform the selections but under-perform relative to the CNN. 

\emph{\bf Further improving performance:}
In figure \ref{fig:roc1} we show further improvements to performance from modifications to our network or training.  Firstly we alter the cross-section based weights applied to the training loss (described in section \ref{sec:CNN}). The background cross-section weights are in some cases are seven orders of magnitude larger than that of the signal sample. This range of values causes some instability for training in some of the deep learning frameworks we use, so we also train the network with applying the natural logarithm of these normalized positive weights. This slightly improves the performance in the false-positive region of interest, though it performs worse than the simple cross-section for higher false-positive rates.

Next, we explore using three input channels in our CNN: namely to separate the energy deposited in the electromagnetic and hadronic calorimeters, and to add the number of tracks reconstructed in the same $\eta/\phi$ bin. Figure \ref{fig:roc1}, shows the further performance improvements. 

Finally a slight further improvement can be seen in figure \ref{fig:roc1} from creating an ensemble of the networks trained with original weights and the log weights (with 3 channels). This is done by taking the mean of the two network outputs as a prediction.

\emph{\bf Robustness to different signal samples and pileup:}
The model described in preceding sections was trained on a specific cascade decay with gluino mass (MGlu) of 1400 GeV, and neutralino mass (MNeu) of 850 GeV. However, in figure \ref{fig:roc4} we show that applying this network to other signal samples with different input masses, without retraining, still offers good performance that exceeds that of the benchmark selections for each sample. 

The above studies used simulated samples without a contribution from additional interactions per beam crossing (pile-up). We also generate samples using the default Delphes ATLAS detector configuration `card' with added pile-up ($\mu = 20$). The 
physics selections benchmark has a higher false-positive rate for this sample (0.001) with a TPR of 0.6 at that point. The CNN still performs well with a TPR of around 0.9 at that FPR. 

\begin{figure*}[!t]
\centering
\begin{minipage}{0.30\textwidth}
\includegraphics[width=\textwidth]{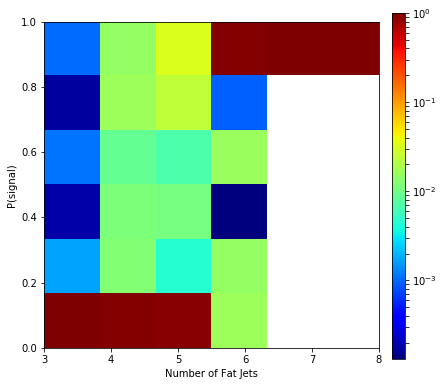}
\caption{Comparison of CNN output with Number of Fat Jets}
\label{fig:jetvars}
\end{minipage}
\hfill
\begin{minipage}{0.30\textwidth}
\includegraphics[width=\textwidth]{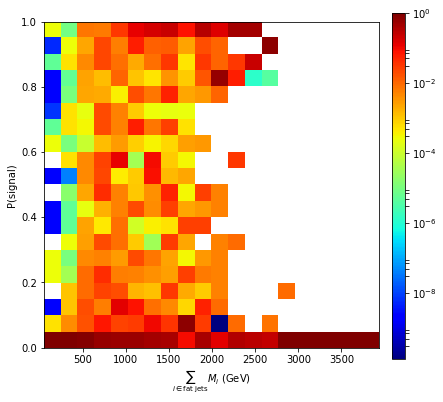}
\caption{Comparison of CNN output with sum of jet mass}
\label{fig:jetvars2}
\end{minipage}
\hfill
\centering
\begin{minipage}{0.34\textwidth}
\includegraphics[width=\textwidth,trim={0.4cm -1.50cm 0.4cm 1.0cm} ,clip]{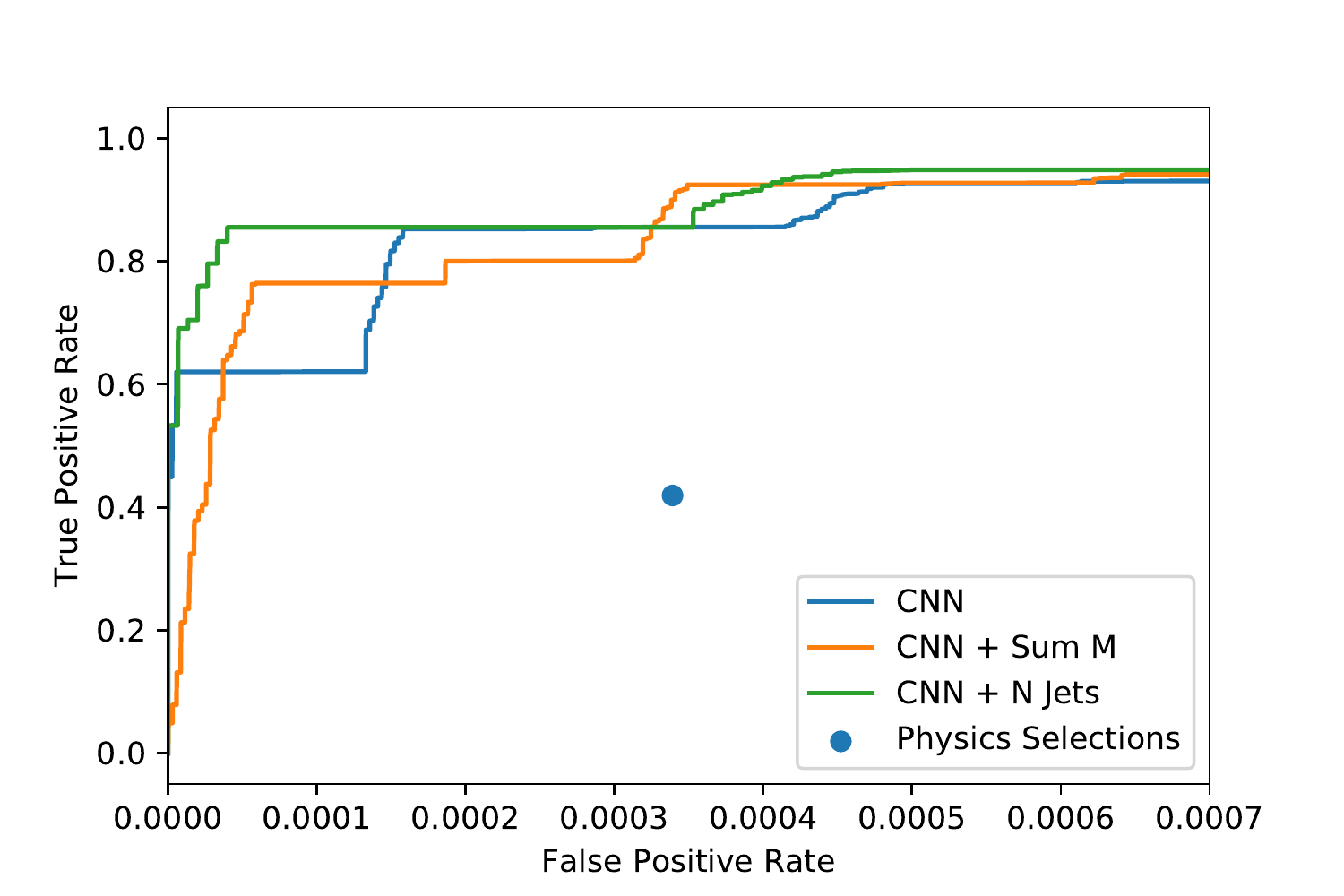}
\caption{Change in performance on combining CNN output with jet variables}
\label{fig:roc6}
\end{minipage}
\end{figure*}

\subsection{Relation between CNN and physics variables}
\label{sec:Results2}
In figures \ref{fig:jetvars} and \ref{fig:jetvars2} we compare the predicted class probability of the CNN with two of the jet variables used in the physics analysis and a clear correlation can be seen. We also train a one-layer network combining the CNN output with each of these physics variables in turn. The performance is shown in figure \ref{fig:roc6} and there is little or no further improvement within the statistics of the sample used, suggesting that the CNN is capturing much of the discriminating power of these jet variables, despite not being explicitly trained on them.

\subsection{Training Time Performance on CPU and GPU architectures}
\label{sec:timing}

\begin{figure*}[!t]
\centering
\includegraphics[width=0.80\textwidth]{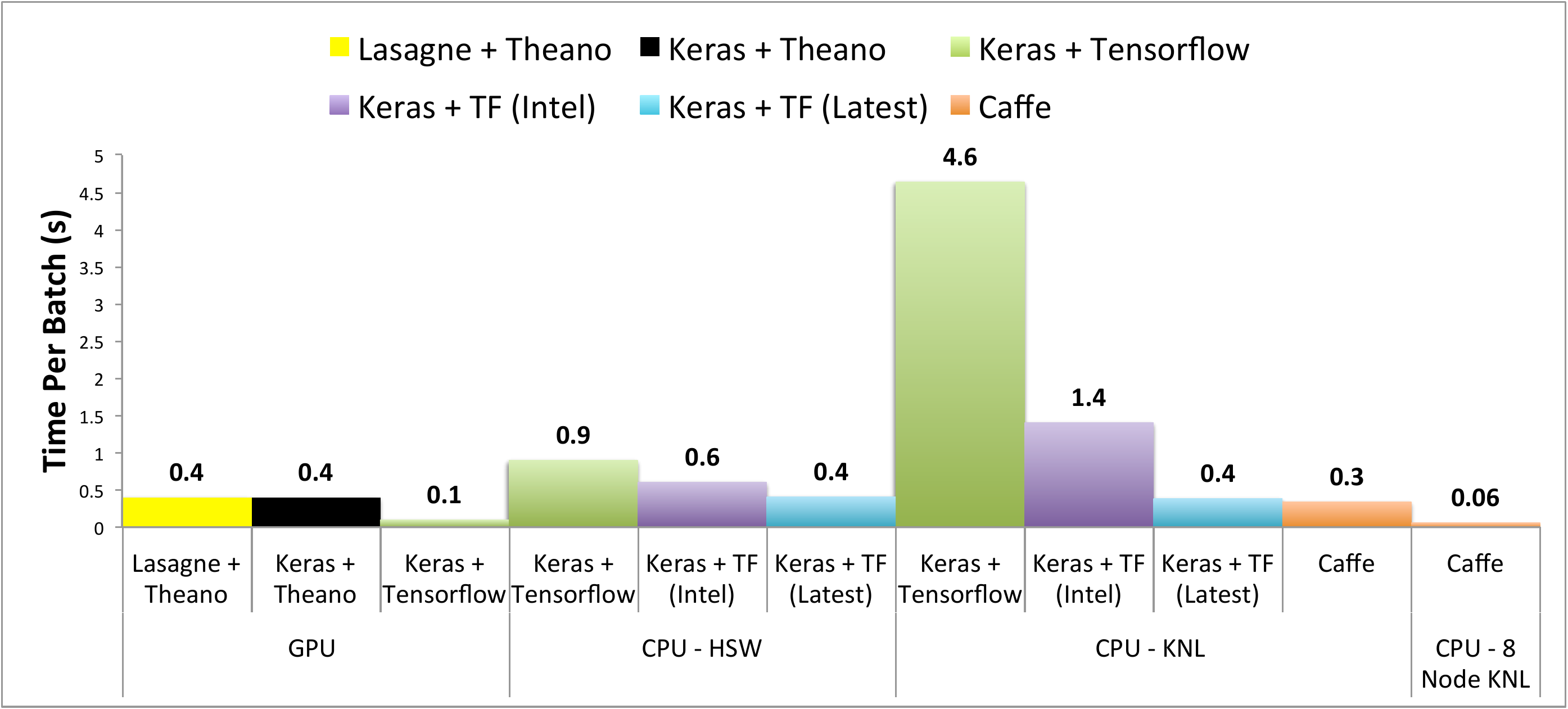}
\caption{Comparison of time per batch (batch size=512) for different deep-learning frameworks on GPU, Haswell (HSW) and Knights Landing (KNL) CPU}
\label{fig:timing}
\end{figure*}


In figure \ref{fig:timing} we compare the training time for implementations of our network in different popular deep-learning frameworks and on different architectures. We report the time per batch of 512 examples excluding I/O. Not all implementations have been optimized so the aim is not to provide detailed comparisons of frameworks but rather to drive improved CPU performance. It can clearly be seen that the default Tensorflow (v1.2) \cite{tfsoft} implementation performed very poorly on the KNL architecture. Several optmizations were recently introduced by Intel for Tensorflow on CPU \cite{IntelTF} building on the Intel Math Kernel Libary (MKL) \cite{MKL2017DeepLearning}, such as multi-threaded convolutional layers, vectorizing over channels or filters and cache blocking. These offer substantial improvements as shown in the times labeled `TF (Intel)' in figure \ref{fig:timing}. Further optimizations have since been carried out in a collaboration between NERSC and Intel using the architecture and data from this paper, as well as other use-cases, resulting in the `TF (Latest)' measurement in figure \ref{fig:timing}. These include MKL implementation of element-wise operations to avoid MKL to Eigen conversions. This version is currently available for use on NERSC systems and will be included into future releases of Tensorflow. Code, datasets and recipes are provided in \ref{sec:code}.

Performance improvements for CPU were also made in IntelCaffe  \cite{intelcaffe}. Figure \ref{fig:timing} shows this implementation also performs well. In addition a multi-node version of IntelCaffe using the Intel MLSL library \cite{mlsl} is available at NERSC. Figure  \ref{fig:timing} shows that using 8 nodes (with a constant batch size per node) one can train this network approximately 6x faster than a single node. With other collaborators, we have scaled this Caffe implementation on the same problem, but with large 224x224 images, to 9600 KNL nodes on Cori. That study is reported in detail in \cite{kurth2017deep}. 


\section{Conclusions}
\label{sec:conclusion}
We have implemented deep convolutional networks on large whole-detector images directly for physics analysis. We find this offers improved sensitivity than physics-variable based selections, and shallow classifiers using those variables, without the need for jet reconstruction. Further improvements come, in particular, from adding multiple sub-detectors as channels. This network is robust to pileup and can be applied to other masses of the signal sample without retraining. 

Furthermore, we have used this implementation and data to benchmark and improve popular deep learning libraries on CPU architectures including XeonPhi/KNL on the NERSC Cori supercomputer as well as to demonstrate distributed training up to 9600 KNL nodes. 

\section*{References}
\bibliographystyle{iopart-num}
\bibliography{references}

\section*{Acknowledgements}
We thank Ben Nachman and Brian Amadio for many valuable discussions and physics input, and Mustafa Mustafa and the Intel Tensorflow team for CPU optimizations. This research used resources of the National Energy Research Scientific Computing Center (NERSC), a DOE Office of Science User Facility supported by the Office of Science of the U.S.
Department of Energy under Contract No. DE-AC02-05CH11231.

\appendix
\section{Benchmark analysis and pre-selection}
\label{tab:presel}
The selections applied for the benchmark analysis were (following \cite{ATLAS:2016nij}):

\begin{tabular}{l}
{\bf Fat-jet object selection:} \\
AntiKt R=1.0 trimmed ($R_{trim}= 0.2$, $P_{Tfrac} = 0.05$)\\
$P_{T} > 200$ GeV , $|\eta| < 2.0$ \\
{\bf Preselection:} \\
Leading Fat-jet $P_T > 440 GeV$ \\
$N_{Fat-Jet} > 2$ \\
{\bf Analysis Selection:} \\
$|\Delta \eta_{12}|$   between leading two Fat-jets $< 1.4$ \\
($N_{Fat-Jet}  >= 4$ AND  $\Sigma M_{Fat-jet} > 800$ GeV) \\
OR ($N_{Fat-Jet} >= 5$ AND  $\Sigma M_{Fat-jet} > 600$ GeV) \\
\end{tabular}

\section{Datasets, Code and Recipes}
\label{sec:code}
Example code for implementing the networks used in this study, together with datasets and recipes for running at NERSC are provided at 
\url{http://www.nersc.gov/users/data-analytics/data-analytics-2/deep-learning/deep-networks-for-hep/}

\end{document}